\documentclass[twoside,11pt]{article}

\usepackage{jmlr2e}

\usepackage[multiple]{footmisc}

\usepackage{listings}
\usepackage{tablefootnote}
\usepackage{amsmath}
\usepackage{multirow}
\usepackage{color}

\newcommand{\tens}[1]{#1}

\usepackage{lastpage}
\jmlrheading{21}{2020}{1-\pageref{LastPage}}{1/18; Revised
2/20}{3/20}{18-008}{Alexander Novikov, Pavel Izmailov, Valentin Khrulkov, Michael Figurnov, and Ivan Oseledets}

\ShortHeadings{Tensor Train Decomposition on TensorFlow (T3F)}{Novikov, Izmailov, Khrulkov, Figurnov, and Oseledets}
\firstpageno{1}

\begin{document}

\title{Tensor Train Decomposition on TensorFlow (T3F)}

\author{\name Alexander Novikov$^{1,2}$ \email sasha.v.novikov@gmail.com \\
\name Pavel Izmailov$^{3}$ \email pi49@cornell.edu \\
\name Valentin Khrulkov$^{4}$ \email valentin.khrulkov@skolkovotech.ru \\
\name Michael Figurnov$^{1}$ \email michael@figurnov.ru\\
\name Ivan Oseledets$^{2,4}$ \email i.oseledets@skoltech.ru \\
\AND
       \addr $^{1}$ National Research University Higher School of Economics,~
       Moscow, Russia\\
       \addr  $^{2}$ Institute of Numerical Mathematics RAS,~
       Moscow, Russia\\
       \addr  $^{3}$ Cornell University,~
       Ithaca NY, USA\\
       \addr  $^{4}$ Skolkovo Institute of Science and Technology,~
       Moscow, Russia
       }

\editor{Alexandre Gramfort}

\maketitle

\begin{abstract}%
Tensor Train decomposition is used across many branches of machine learning. We present T3F---a library for Tensor Train decomposition based on TensorFlow. T3F supports GPU execution, batch processing, automatic differentiation, and versatile functionality for the Riemannian optimization framework, which takes into account the underlying manifold structure to construct efficient optimization methods. The library makes it easier to implement machine learning papers that rely on the Tensor Train decomposition. T3F includes documentation, examples and $94\%$ test coverage.
\end{abstract}

\begin{keywords}
  tensor decomposition, tensor train, software, gpu, tensorflow
\end{keywords}

\section{Introduction}

Methods based on tensor decompositions have become ubiquitous in the machine learning community. They are used for analyzing theoretical properties of deep networks~\citep{cohen2016expressive, cohen2016convolutional, khrulkov2017expressive,janzamin2015beating}, network compression~\citep{lebedev2014speeding,novikov2015tensorizing,yu2017long}, training probabilistic models~\citep{anandkumar12spectral, jernite13, song13}, parametrizing recommender systems~\citep{frolov2017tensor}, and many more. In this work, we present a library for a particular tensor decomposition, namely, the Tensor Train decomposition~\citep{oseledets2011ttMain}.

The Tensor Train (TT) format~\citep{oseledets2011ttMain} (which is also called the Matrix Product State (MPS) format in the physics community~\citep{fannes-mps-1992}) is a generalization of matrix low-rank format to higher-dimensional tensors.
A tensor $\tens{A} \in \mathbb{R}^{n^d}$ is said to be represented in the TT-format if each of its elements can be represented as a product of factors
\begin{equation*}
    A_{i_1 \ldots i_d} = \sum_{\alpha_1=1}^{r} \ldots \sum_{\alpha_{d-1}=1}^{r}G^1_{i_1 \alpha_1}G^2_{\alpha_1 i_2 \alpha_2}G^3_{\alpha_2 i_3 \alpha_3} \ldots G^d_{\alpha_{d-1} i_d},
\end{equation*}
where $G^1 \in \mathbb{R}^{n \times r}$, $G^2, \ldots, G^{d-1} \in \mathbb{R}^{r \times n \times r}$, and $G^d \in \mathbb{R}^{r \times n}$ are called \emph{TT-cores}.
The number of values $r$ the auxiliary indices $\alpha_k$ can take is called \emph{TT-rank}. It controls the trade-off between memory/computational efficiency and the representational power of the format. Representing tensor $\tens{A}$ via explicit enumeration of its elements requires $n^d$ scalars, while representing it in the TT-format only requires $rn (r (d - 2) + 2)$.

The Tensor Train decomposition is supported by many libraries (see Section~\ref{sec:other-libraries}), but some of the recent papers that use it for machine learning purposes had to rewrite the core functionality from scratch. We believe that this is due to the lack of a library with simultaneous support of the following features important for machine learning research: GPU execution, automatic differentiation, parallel processing of a batch of tensors, and advanced support for Riemannian geometry operations---a technique for speeding up the optimization when the parameters are constrained to have a compact Tensor Train (TT) representation (or other constraint sets that form a smooth manifold).

In the presented library, we aim to make the results of the recent machine learning papers utilizing the TT-format easy to reproduce and provide a flexible framework for developing new ideas.
The library is released\footnote{\url{https://github.com/Bihaqo/t3f}} under MIT license and is distributed as a PyPI package\footnote{\url{https://pypi.python.org/pypi/t3f}} to simplify the installation process. The documentation is also available online\footnote{\url{https://t3f.readthedocs.io}}. The library includes Jupyter notebook examples, e.g. for performing tensor completion by assuming that the result has a low TT-rank and minimizing the residual between the tensor and the known elements by gradient descent or by Riemannian optimization. The library supports both graph and eager TensorFlow execution, making it easy to prototype, debug and deploy the resulting models. The library has $94\%$ test coverage.

\section{Related Work \label{sec:other-libraries}}
Many libraries implement operations for working with the TT-format. The most similar ones to the presented library (T3F) are tntorch~\citep{tntorch} and TensorLy~\citep{kossaifi2019tensorly}. The main differences between T3F and these libraries are that tntorch uses PyTorch as a backend, T3F uses TensorFlow, and TensorLy supports PyTorch, TensorFlow and MxNet as backends. Another difference is that neither tntorch nor TensorLy support Riemannian optimization operations.

Other libraries include: ttpy~\citep{ttpy}, mpnum~\citep{Suess2017}, scikit\_tt ~\citep{scikittt}, mpys~\citep{mpys}, OSMPS~\citep{jaschke2018open},  evoMPS~\citep{evoMPS}, TT-Toolbox~\citep{TT-Toolbox}, ITensor~\citep{itensor}, libtt~\citep{libtt} and MPS~\citep{mps}. Unlike these libraries, T3F supports GPU execution, automatic differentiation (including Riemannian autodiff) and batch processing. However, some of these libraries support advanced algorithms such as DMRG~\citep{khoromskij2010dmrg} or AMen~\citep{dolgov2014alternating}, which T3F does not.

\section{Implementation Details}
The library provides two base classes: \texttt{TensorTrain} and \texttt{TensorTrainBatch} that support storing one tensor in the Tensor Train format and a batch of such tensors respectively, i.e. a list of tensors of the same shape that are supposed to be processed together. These two classes support most of the API of \texttt{tf.Tensor} class (e.g. \texttt{.op}, \texttt{.name}, and \texttt{.get\_shape()}). Under the hood, these classes are containers for the TT-cores represented as \texttt{tf.Tensor} objects, plus  lightweight meta-information. This design allows one to work with the TT-cores directly if necessary. The rest of the library is a collection of functions, each taking as an input one or two TT-objects and outputting a TT-object or a numerical value depending on the semantics of the function. For example, the function \texttt{t3f.multiply(left, right)} implements elementwise multiplication of two TT-tensors (or batches of TT-tensors) but also supports multiplication of a TT-tensor by a number. As an output, this function returns a \texttt{TensorTrain} or a \texttt{TensorTrainBatch} object.

A typical machine learning use case would be to initialize a random TT-matrix and use it to define a neural network layer (using \texttt{t3f.matmul(tt\_matrix, x)} to compute the matrix-by-vector product instead of the usual \texttt{tf.matmul(matrix, x)}). Then, one can use the optimization functionality of TensorFlow to train the model with respect to its parameters including the TT-cores of the TT-matrix~\citep{novikov2015tensorizing,NIPS2016_6211,lebedev2014speeding,yang2017tensor}. See this and other example in the documentation\footnote{\url{https://t3f.readthedocs.io/en/latest/tutorials/tensor_nets.html}}.

A typical non-machine-learning use case would be finding the solution of a PDE discretized on a grid in the TT-format~\citep{grasedyck2013literature,khoromskij2018tensor}. One would initialize a tensor in the TT-format and then iteratively perform the updates as a sequence of operations on the tensor. At the end of each iteration, one would typically use the rounding operation \texttt{t3f.round(iterate, max\_tt\_rank=r)}, which finds the best approximation to a TT-tensor among TT-tensors of lower TT-rank, in order to restrict the TT-rank growth. T3F is a good fit for these applications because of the Riemannian optimization and GPU support and it was already used by~\cite{rakhuba2019low}.

\subsection{Batch Processing}
Most operations accept broadcasting and getting a batch of TT-objects as an input. For example, \texttt{C = t3f.matmul(A, B)} for a batch of TT-matrices \texttt{A} and a TT-matrix \texttt{B} will return a batch of TT-matrices \texttt{C} where $\texttt{C}_i = \texttt{A}_i 
\texttt{B}$ and the result is computed in parallel across the batch dimension. This functionality is important for efficient mini-batch processing of large scale data sets.

\subsection{Riemannian Geometry Operations}
One of the advantages of the Tensor Train format is that the set of tensors of fixed TT-rank forms a Riemannian manifold, which allows the use of Riemannian geometry ideas to speed up tensor calculus while preserving theoretical guaranties (see~\cite{steinlechner2016riemannian} for more details). The T3F library has rich support for Riemannian operations, the most basic being projecting a TT-object $\tens{A}$ (or a batch of them) onto the tangent space of another TT-object~$\tens{B}$. We denote this projection operation by $P_{\tens{B}} \tens{A}$.

Other supported operations are special cases of combining this basic projection operation with non-Riemannian operations but are heavily optimized by exploiting the structure of objects that are projected onto the same tangent space. Such operations include projecting a weighted sum of a batch of TT-objects on a tangent space (used for efficiently computing the Riemannian gradient on a mini-batch of objects).
Mathematically, this function implements projecting a weighted sum of tensors $\tens{S} = P_{\tens{B}} \left ( \sum_i c_i \tens{A}_i \right )$. The same operation can be implemented by a summation followed by a projection operation in asymptotic complexity $\mathcal{O}(b d r_A r_B n (r_A b + r_B))$, where $b$ is the batch-size and $r_A$ and $r_B$ are the TT-ranks of the tensors $\tens{A}$ and $\tens{B}$. But the tailored \texttt{project\_sum} operation requires only $\mathcal{O}(b d  r_A r_B n (r_A + r_B))$ for the same operation. The idea behind \texttt{project\_sum} is to use the linearity of the projection $\tens{S} = \sum_i c_i P_{\tens{B}} \tens{A}_i$, and the fact that elements of the same tangent space can be added efficiently.
Other tailored operations include computing the Gram matrix of a batch of tensors from the same tangent space and projecting the matrix-by-vector product onto a tangent space $P_{b} ({A} {c})$.

Additionally, T3F is the only library that supports automatic Riemannian differentiation for TT-format, i.e. given a function $f(\tens{X})$ the library can automatically (and with optimal asymptotic complexity) find the projection of the gradient of the function on the tangent space of the tensor $\tens{X}$: $P_{\tens{X}} \frac{\partial f}{\partial \tens{X}}$. The product $P_{T_{\tens{X}}} \left(\nabla^2 f(\tens{X}) \right) P_{T_{\tens{X}}}\tens{Z}$ of the linearized Riemannian Hessian by a given tensor $\tens{Z}$ is also supported.

\section{Benchmarking}
We benchmark the basic functionality of T3F on CPU and GPU and compare its performance against an alternative library TTPY. To reproduce the benchmark on your hardware, see \texttt{docs/benchmark} folder in the T3F library.

For benchmarking, we generated a batch of 100 random TT-matrices of sizes $10^{10} \times 10^{10}$ (so $d=10$ and the TT-representation consists of 10 TT-cores) of TT-rank 10 and a batch of  100 random TT-vectors of size $10^{10} \times 1$. We benchmarked the matrix-by-vector multiplication (`matvec'), matrix-by-matrix multiplication (`matmul'), computing the Frobenius norm (`norm'), and computing the Gram matrix of 1 or 100 TT-vectors. The results are reported in Tbl.~\ref{tbl:t3f-benchmark}. We report that T3F is faster than TTPY for most operation and that batch processing and GPU acceleration yield significant speedups for some operations (Tbl.~\ref{tbl:t3f-benchmark}). Note that TTPY lacks GPU and batch processing support.
\begin{center}
\begin{table}%
  \begin{tabular}{  l c c  c  c c}
    \hline
    \multirow{3}{2cm}{\centering  Op } & \multirow{2}{2.2cm}{\centering TTPY \\ 1 object CPU} & \multirow{2}{2.2cm}{\centering T3F \\ 1 object CPU } & \multirow{2}{2.2cm}{\centering T3F \\ 1 object GPU } & \multirow{2}{2.2cm}{\centering T3F \\ 100 objects CPU } & \multirow{2}{2cm}{\centering T3F \\ 100 objects GPU }  \\[1cm] \hline
matvec          &            11.142 &       1.190 &       0.744 &         1.885 &         0.140 \\
matmul          &            86.191 &       9.849 &       0.950 &        17.483 &         1.461 \\
norm            &             3.790 &       2.136 &       1.019 &         0.253 &         0.044 \\
gram            &             0.145 &       0.606 &       0.973 &         0.021 &         0.001 \\
project &           116.868 &       3.001 &      13.239 &         1.645 &         0.226 \\
    \hline
  \end{tabular}
  \caption{Time in ms for different operations implemented in TTPY library vs T3F on CPU and GPU. The timing for a batch of 100 objects is reported per single object. The comparison is made on an NVIDIA DGX-1 station with Tesla V100 GPUs (using only 1 GPU at a time) in double precision. \label{tbl:t3f-benchmark}}
\end{table}
\end{center}

\acks{}
This work was partially funded by the Ministry of Science and Education of Russian Federation as a part of Mega Grant Research Project 14.756.31.0001

\newpage

\vskip 0.2in

\bibliography{t3f}

\end{document}